\documentclass[submission,copyright,creativecommons]{eptcs}
\providecommand{\event}{14th International Workshop on Foundations of
Coordination\\
Languages and Self-Adaptive Systems (FOCLASA 2015)} 
\usepackage{breakurl}             
\usepackage{color}
\usepackage{enumerate}
\usepackage{amsfonts}
\usepackage{amsmath,amsthm}
\usepackage{amssymb}
\usepackage{latexsym}
\usepackage{soul}
\usepackage[pdftex]{graphicx}

\title{Service Choreography, SBVR, and Time\thanks{This work was partly funded by the UK Research Council EPSRC, under the project
{\it Evolution and Resilience of Industrial Ecosystems} (ERIE)
, Contract No. EP/H021779/1.}}
\author{Nurulhuda A. Manaf, Sotiris Moschoyiannis, and Paul J. Krause
\institute{Department of Computer Science, University of Surrey\\
Guildford, Surrey, GU2 7XH, UK}
\email{\{n.amanaf, s.moschoyiannis, p.krause\}@surrey.ac.uk}
}

\def\titlerunning{Service Choreography, SBVR, and Time}
\def\authorrunning{N.A. Manaf, S. Moschoyiannis  \& P.J. Krause}
\begin{document}
\maketitle

\begin{abstract}
We propose the use of structured natural language (English) in specifying service choreographies, focusing on the {\it what} rather than the {\it how} of the required coordination of participant services in realising a business application scenario. The declarative approach we propose uses the OMG standard {\it Semantics of Business Vocabulary and Rules} (SBVR) as a modelling language. The {\it service choreography} approach has been proposed for describing the global orderings of the invocations on interfaces of participant services. We therefore extend SBVR with a notion of time which can capture the coordination of the participant services, in terms of the observable message exchanges between them. The extension is done using existing modelling constructs in SBVR, and hence respects the standard specification. The idea is that users - domain specialists rather than implementation specialists - can verify the requested service composition by directly reading the structured English used by SBVR. At the same time, the SBVR model can be represented in formal logic so it can be parsed and executed by a machine.
\end{abstract}

\section{Introduction}
\label{sec:intro}

There is increasing interest in developing distributed applications that involve stand-alone services from different organisations on the web. However, the coordination of the interactions between the underlying services in building such applications remains a challenge. Sustained efforts by the web services community have culminated in the \emph{service choreography} approach \cite{WS02} which is concerned with describing the conversation across different participating services (global perspective) as well as the \emph{service orchestration} approach \cite{WS03} which describes the interaction scenario from an individual service'€™s viewpoint. Service choreography in particular, is intended to capture the coordination of the participant services, in terms of the observable message exchanges between them. This is given mostly in terms of the orderings of these interactions during the execution of the corresponding business activity.

The orderings of the interactions (invocations on interfaces) between the underlying services is key in coordination as they capture the dependencies between participant services and thus the correctness of the application design. Well known issues (e.g., see \cite{M02}) that involve the orderings of interactions include \emph{deadlock} and \emph{race conditions} (a situation where two or more messages are competing to arrive first, so while the appear to be ordered in a given execution they are effectively unordered). In the context of service choreography, verification additionally comes in the form of  \emph{conformance} and \emph{realisation} \cite{Su07, Bultan07}. Moreover, if choreographies are to be equipped with transactional guarantees \cite{S01}, meaning that a series of compensations are performed upon failure, the ordering of the interactions is doubly important.

Declarative approaches in the Business Rules realm \cite{Ros03}, \cite{Dat04} focus on \emph{what} rather than \emph{how}. The 'what' and the 'how' of a solution to a computing problem are quite different. The 'what' refers to the properties of a solution whereas the 'how' refers to the steps followed to achieve the solution. Declarative programming focuses on specifying the 'what' and using a general-purpose engine for reaching the goal. An example declarative language is SQL, which specifies properties of data but not the way to retrieve it. The latter is left to the database management system (DBMS) implementation \cite{Dat00}. Imperative programming focuses on the 'how', bypassing the need to define the properties of the required solution since programmers can guarantee the desired properties by directly controlling the algorithm. Java and C are generally considered imperative languages.

With respect to specifying service choreography, the declarative approach can express business requirements intuitively, e.g., see \cite{Dat04, F09}. The Business Rules manifesto \cite{Ros03} builds the business requirements on the premise that rules or policies in a business application scenario should be expressed declaratively in natural-language sentences for the business audience. A rule is distinct from any enforcement defined for it. A rule and its enforcement are separate concerns. Also, rules apply across processes and procedures. In addition, the issue of understandability in expressing and modelling complex business requirements is important especially if we want the provision for the domains specialists to validate the specification against their business models. Note that domains specialists (business analysts, stakeholders) rather than implementation specialists are best positioned to validate against the business activities taking place in practice. In our approach, the choreography specification is expressed in terms of statements like the following:
\begin{center}
\textcolor[rgb]{1.00,0.50,0.00}{It is obligatory that} \textcolor[rgb]{1.00,0.50,0.00}{each}  \textcolor[rgb]{0.00,0.59,0.00}{\ul{\textbf{rental car}}}  \textcolor[rgb]{0.00,0.07,1.00} {\emph{is owned by}}   \textcolor[rgb]{1.00,0.50,0.00}{at least one}  \textcolor[rgb]{0.00,0.59,0.00}{\ul{\textbf{branch}}}
\end{center}
In addition, a declarative approach typically starts with an unconstrained view of the specification and gradually constrains it, by means of adding rules, as the intended behaviour of the service choreography becomes more clear. This is in contrast to the more traditional imperative approach, which tends to be more restrictive and sometimes results in introducing artificial decision points, or forcing premature decision points, for the practitioner or over- / under-specifying what actually happens \cite{M10}.

Work on formal semantics in this area has focused more on the imperative (or procedural) approach and service orchestration and less so on the declarative approach and service choreography. Existing work, e.g., \cite{M10, RB01,Sc02} that takes a declarative approach tends to focus on reasoning about consistency of the rule set, which of course is an important aspect of verification, but have not looked into explicitly capturing the orderings, in terms of observable message exchanges in a choreography.  In addition, and to the best of our knowledge, none of the current proposals for a declarative approach to service choreography has attempted to provide the end-user with something close to natural language.

In this paper, we propose a declarative approach which builds on using the \emph{Semantics of Business Vocabulary and Rules} (SBVR) \cite{WS04}  for the specification of service choreographies. SBVR is a standard maintained by the Object Management Group (OMG) and uses structured natural language, which makes it specifically understandable by humans. The business rule given earlier is actually written in SBVR and makes use of two Fact Types (cf Section 2). There is no explicit notion of time in SBVR. In order to capture the global ordering constraints on observable actions (invocations) in a service choreography we describe the use of sequencing of Fact Types in an SBVR model, together with the modelling constructs of \emph{objectification} and \emph{actuality}. This allows us to specify, for example, that €a product is received by the customer only after it is delivered by the shop€™.

The remainder of this paper is structured as follows. Section 2 contains a brief account of the business rules approach and SBVR. In Section 3, we build an SBVR model for a case study, an Online Photo Shop, which focuses on the ordering of service interactions. This forces us to look at temporal aspects and thus Section 4 describes our handling of time ordering within SBVR. Sections 5 gives a brief account of related work. Some concluding remarks and ideas for future work are included in Section 6.

\section{ The Business Rules approach and SBVR}
\label{sec:SBVR}



Several specification or modelling languages for specifying interactions between services in a business application scenario are available to practitioners, with varying levels of adoption, such as the \emph{Business Process Model Notation} (BPMN) \cite{WS05}, \emph{Web Services Choreography Description Language} (WS-CDL) \cite{WS02}, \emph{Web Services Business Process Execution Language} (WS-BPEL) \cite{WS03}. These languages require training to read and write and hence may not lend themselves naturally to be used by the end-user directly.

The OMG standard {\it Semantics of Business Vocabulary and Rules} (SBVR) \cite{WS04} is gaining ground as a basis for system specification. By inception, SBVR is intended to provide a way to capture specifications in natural language and represent them in formal logic so they can be machine processed. Users are able to verify the specification directly by reading the structured natural language used by SBVR which can then be parsed and executed by a machine. In line with the Business Rules Approach \cite{Ros03}, it follows the doctrine: "{\it Rules build on facts, and facts build on concepts as expressed by terms. Terms express business concepts; facts make assertions about these concepts; rules constrain and support these facts.}"

\begin{figure}[!hbtp]
  \centering
    \fbox{\includegraphics[width=0.985\textwidth]{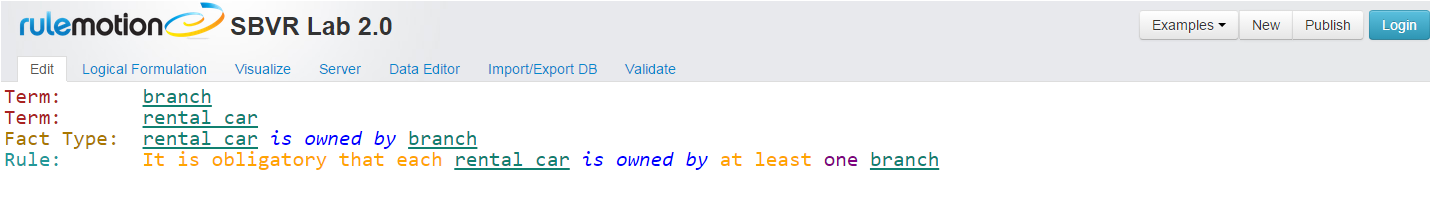}}
  \caption{Part of an SBVR model for the specification of the Rental Car case study}\label{rspec}
\end{figure}

As argued in \cite{MaKRuleML09}, while SBVR is a meta-model with models natively expressed as logical formulations, its most common serialization is SBVR Structured English. Terms (e.g., \textcolor[rgb]{0.00,0.59,0.00}{\ul{\textbf{branch}}}), Fact Types (e.g., \textcolor[rgb]{0.00,0.59,0.00}{\ul{\textbf{rental car}}} \textcolor[rgb]{0.00,0.07,1.00}{\emph{is owned by}} \textcolor[rgb]{0.00,0.59,0.00}{\ul{\textbf{branch}}}), and rules (e.g. ˜\textcolor[rgb]{1.00,0.50,0.00}{It is obligatory that} \textcolor[rgb]{1.00,0.50,0.00}{each}  \textcolor[rgb]{0.00,0.59,0.00}{\ul{\textbf{rental car}}}  \textcolor[rgb]{0.00,0.07,1.00} {\emph{is owned by}}   \textcolor[rgb]{1.00,0.50,0.00}{at least one}  \textcolor[rgb]{0.00,0.59,0.00}{\ul{\textbf{branch}}}) are combined into models.
An example of an SBVR model can be seen in Figure \ref{rspec}. It refers to the Rental Car case study included in its spec document\cite{WS04}.

The rule in Figure \ref{rspec} is written using our web-based SBVR editor \cite{SBVReditor} maintained by Rulemotion\footnote{With thanks to Rulemotion, the editor {\it SBVR Lab 2.0} is available at \url{http://sbvr.co}} and is a representation of higher-level facts that use the deontic constraint, \textcolor[rgb]{1.00,0.50,0.00}{obligatory} on the constraint defined by the rule. The quantifications, \textcolor[rgb]{1.00,0.50,0.00}{each} and \textcolor[rgb]{1.00,0.50,0.00}{at least one} show the restriction of rental car belonging. Furthermore, "\textcolor[rgb]{0.00,0.07,1.00}{\emph{is owned by}}' is the designation for the Fact Type in Table 1. Fact Type is constructed based on identified Terms (a noun concept, \textcolor[rgb]{0.00,0.59,0.00}{\ul{\textbf{rental car}}} and \textcolor[rgb]{0.00,0.59,0.00}{\ul{\textbf{branch}}}). Thus, the combination of deontic constraint, quantifier, terms and fact type will yield a constructive rule. This type of rule can be used by domain specialists (e.g., business analysts) in defining the business model or activity to be performed with a choreography.

\cite{D01} provides a syntax and semantics for the logical formulations of SBVR, a \emph{first-order deontic-alethic logic (FODAL)}. It is an extension of first-order logic which is a combination of standard-deontic logic (SDL) and normal modal logic, S4. The syntax of \emph{FODAL} \cite{D01} includes a set of propositional connectives ($\neg$, $\wedge$), a universal quantifier ($\forall$), an infinite set $\mathcal{P}$ (predicate symbols), an infinite set $\mathcal{V}$ (variable symbols), and, modal operators ($\Box$ (necessity) and \textbf{\emph{O}} (obligation)) for alethic and deontic respectively. To formalise SBVR rules, \emph{FODAL} follows the first-order modal formulae which are specified by the rules :
\begin{itemize}
  \item Every atomic formula is a formula.
  \item If \emph{X} is a formula, so is $\neg{\emph{X}}$.
  \item If \emph{X} and \emph{Y} are formulas, then \emph{X}$\wedge$\emph{Y} is a formula.
  \item If \emph{X} is a formula, so are $\Box{X}$ and \emph{\textbf{O}X}.
  \item If \emph{X} is a formula and $\nu$ is a variable, then $\forall{\nu}$\emph{X} is a formula.
\end{itemize}
The usual definition is used for the existential quantifier ($\exists$) and other propositional connectives ($\vee$, $\rightarrow$, $\leftrightarrow$). However, there are additional modal operators defined for possibility ($\diamondsuit$), permission (\emph{\textbf{P}}), and prohibition (\textbf{\emph{F}}) which are "It is possible that $\phi$" is logically equivalent to "It is not necessary that not $\phi$" ($\diamondsuit\phi\equiv\neg\Box\neg\phi$), "It is permitted that $\phi$" is logically equivalent to "It is not obligatory that not $\phi$" (\textbf{\emph{P}}$\phi\equiv\neg$\textbf{\emph{O}}$\neg\phi$), and "It is forbidden that $\phi$" is logically equivalent to "It is obligatory that not $\phi$" (\textbf{\emph{F}}$\phi\equiv$\textbf{\emph{O}}$\neg\phi$).

In addition, \cite{D01} provides a \emph{Kripke} semantics for \emph{FODAL} as well as the proofs of its sound and complete axiomatisations with respect to the semantics \cite{mastersthesis}. The axioms of \emph{FODAL} implies the combination of the axiom systems for the propositional modal logics \textbf{\emph{S4}} and a serial relation of a deontic modality behaviour (\textbf{\emph{KD}}) as well as the interaction between alethic and deontic modalities. The \emph{FODAL} axioms as in \cite{D01} are shown as follows:
\begin{center}
\begin{description}
  \item[(Tautologies \textbf{\emph{S4}})] Any FOL substitute-instance of a theorem of \textbf{\emph{S4}}
  \item[(TAutologies \textbf{\emph{KD}}] Any FOL substitute-instance of a theorem of \textbf{\emph{KD}}
  \item[(Vacuous $\forall$)] $\forall{x}\phi\equiv\phi$, provided $x$ is not free in $\phi$
  \item[($\forall$ Distributivity)] $\forall{x}(\phi\rightarrow\psi)\rightarrow(\forall{x\phi}\rightarrow\forall{x\psi}$)
  \item[($\forall$ Permutation)] $\forall{x}\forall{y\phi}\rightarrow\forall{y}\forall{x\phi}$
  \item[($\forall$ Elimination)] $\forall{y}(\forall{x\phi(x)}\rightarrow\phi(y))$
  \item[Necessary \textbf{\emph{O}}] $\Box\phi\rightarrow$\textbf{\emph{O}}$\phi$
\end{description}
\end{center}
Even though \emph{FODAL} is undecidable, \cite{D01} identifies a decidable fragment  of \emph{FODAL} logic. This is the set of atomic modal sentences with at most two variables, all predicate symbols with at most unary, and the set of atomic modal sentences in which are applied to subformulas from the guarded fragment of firs-order logic.

In the context of service choreography, it is important to capture the ordering of observable actions (service interactions) as discussed before. SBVR does not include a notion of time and therefore with respect to time and ordering, OMG has supplemented it with the \emph{Date-Time Vocabulary} (DTV) \cite{WS08}. To be more precise, DTV expresses the specification in the form specified in Annex C of SBVR as defined by OMG \cite{WS04}. It was introduced to encapsulate the SBVR rules that involve concepts such as date and time (excluding real-time processing) which are frequently used in everyday business activities across a wide range of business scenarios.

Two types of time are considered in DTV. Type 1 refers to a time period, an explicit time interval. Type 2 uses temporal concepts to define a relationship between \emph{situation kinds} and \emph{occurrences}. These are used to represent the potential real activities or events that occur multiple times in a business environment.
In our work on choreography specification we apply Type 2 from DTV as well as the SBVR verb concept \emph{objectification} \cite{WS04} in order to express the ordering of exchanged messages and corresponding Fact Types, e.g., €™\emph{\textcolor[rgb]{0.00,0.59,0.00}{\ul{\textbf{A}}} \textcolor[rgb]{0.00,0.07,1.00} {\emph{before}} \textcolor[rgb]{0.00,0.59,0.00}{\ul{\textbf{B}}}}€™ . This will be further discussed in Section 4.

\section{Service Choreography Specification using SBVR}
\label{sec:sbvr-choreo}

In this section we describe how an SBVR model can be built for the service choreography involved in the Online Photo Shop case study, which was originally studied in \cite{M10}. We identify the need for expressing the ordering relationship between observable events and propose a way to express such orderings in an SBVR model.

\subsection{Online Photo Shop: a case study}\
\label{sec:casestudy}

We look at the case study of an Online Photo Shop in \cite{M10} which provides services for placing orders and printing photographs (and other products) to customers. The business scenario involves a multi-party conversation between several services; \emph{Customer, Photo-Shop, Order, Print,} and \emph{Deliver}. All services are provided by the Online Photo Shop entity, hence in this case they all belong to one organisation. The conversation respects the policies underlining the business activities involved, as described in \cite{M10}.

However, we notice that there are certain problems with the specification of the Online Photo Shop as given in \cite{M10}, such as ambiguities in defining an activity (e.g., 'open order' and 'register') while some prescribed orderings on activities are questionable (e.g., customer may 'pay' before or after Photo Shop performs 'charge'). We have amended the specification slightly to steer away from such issues. This will allow us to focus on using SBVR to specify the choreography rather than elaborating the specification itself. Below is the description of responsibilities by the web services \emph{Customer, Photo-Shop,} and \emph{Order}.
\begin{description}
  \item[\emph{Service Customer}] \emph{provides a service for customer to:}
  \begin{itemize}
    \item \emph{register an account at photo shop by entering data, such as name, address, credit card number and preferred way of delivery through activity "register";}
    \item \emph{pay for ordered products via activity "pay for";}
    \item \emph{receive ordered products via activity "receive".}
  \end{itemize}
  \item[\emph{Service Order}] \emph{allows the customer to order photos and posters via activity "photo" and "poster" respectively by uploading files and selecting wanted formats or to order photo albums by selecting the preference photo album via activity "album".}
  \item[\emph{Service Photo-Shop}] \emph{provides:}
  \begin{itemize}
    \item \emph{the products; photos, posters, and albums. It ensures the shop records the customer's data via activity "update";}
    \item \emph{a service to print ordered photos and posters via activity "print";}
    \item \emph{a service to deliver products (photo, poster, or album) to customer via activity "deliver".}
  \end{itemize}
\end{description}

\begin{table}[!hbtp]
\begin{tabular}{|c|}
  \hline
  \begin{minipage}{6in}
  \vskip 4pt
      \begin{enumerate}
      \setlength{\itemsep}{0pt}
      \setlength{\parskip}{0pt}
      \setlength{\parsep}{0pt}
   \item The shop will not "update" the customer's data before the customer executes activity "register". When the customer executes activity "register", the shop will update its data via activity "update".
   \item After the customer orders photos and posters via activity "photo" and "poster" respectively, the shop prints ordered products via activity "print".
   \item Each ordered product (photo, poster or album) through activity "photo", "poster", or "album" has to be delivered via activity "deliver". The shop will not "deliver" before at least one product is ordered.
   \item Customer can "receive" products only after the shop executes "deliver". All ordered products must be received by the customer through activity "deliver".
   \item Customer has to "pay for" each ordered product (photo, poster or album) made by the customer.
   \item Customer has to "pay for" each ordered product before the shop delivers the ordered products via activity "deliver".
 \end{enumerate}
 \vskip 4pt
 \end{minipage}
 \\
 \hline
 \end{tabular}
  \caption{Amended global constraints for the Online Photo Shop}\label{table:amended constraints}
\end{table}

As a flavour of the kind of changes we made to the case study presented in \cite{M10}, the activity "update" has been introduced in place of "open order" in the 1st constraint shown in Table \ref{table:amended constraints} so as to maintain the semantic meaning of the activities.  The 5th constraint is needed to bind the product that is ordered by the customer to said customer, for each payment. Also, the "pay for" activity is now prescribed to take place before the execution of the "deliver" activity.

\subsection{SBVR model for the Online Photo Shop service choreography}

\begin{figure}[!hbtp]
  \begin{center}
    \fbox{\includegraphics[width=0.985\textwidth]{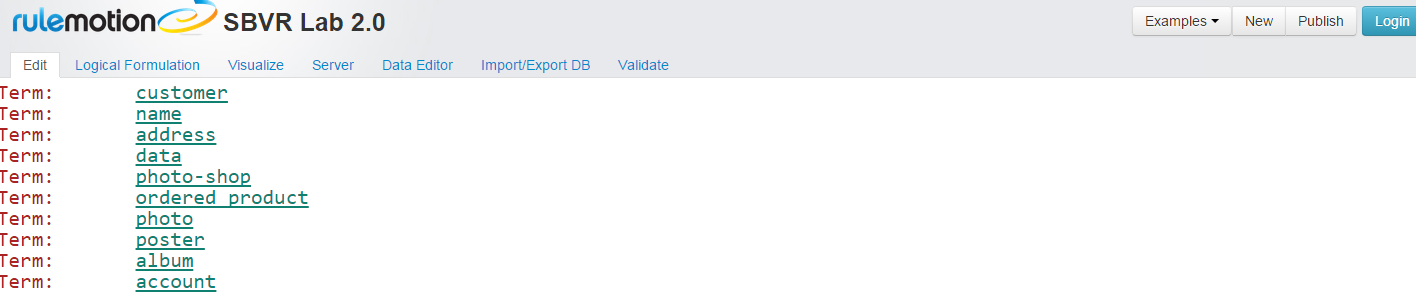}}
  \caption{Terms in the Online Photo Shop case study}
  \label{fig:Terms}
  \end{center}
\end{figure}

\begin{figure}[!hbtp]
  \begin{center}
    \fbox{\includegraphics[width=0.985\textwidth]{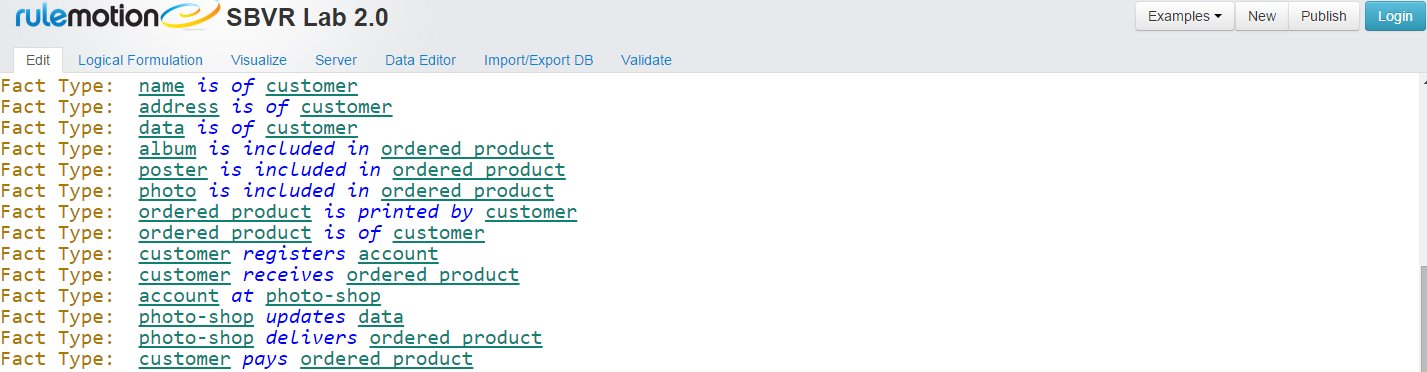}}
  \caption{Fact Types in the Online Photo Shop case study}
  \label{fig:FactTypes}
  \end{center}
\end{figure}

The informal specification of the business scenario can be addressed in a way similar to how entities, attributes and relationships are drawn from a textual specification as done in information modelling and relational database design \cite{Dat04}. Hence, nouns are candidate terms and verbs connecting these nouns together are candidate (binary) fact types. Using this as a rule of thumb, the following Terms (Figure \ref{fig:Terms}) and Fact Types (Figure \ref{fig:FactTypes}) have been extracted for the Online Photo Shop (Section \ref{sec:casestudy}).

As discussed in Section \ref{sec:SBVR}, Term as a noun concept is applied to Fact Type to show the concept that is the meaning of the noun. Fact Type is a combination of one, two or more Terms. The Fact Types formed for the Online Photo Shop case study are shown in Figure \ref{fig:FactTypes}.
For example, the Fact Type, '\textcolor[rgb]{0.00,0.59,0.00}{\ul{\textbf{customer}}} \textcolor[rgb]{0.00,0.07,1.00}{\emph{receives}} \textcolor[rgb]{0.00,0.59,0.00}{\ul{\textbf{ordered product}}}' shows \textcolor[rgb]{0.00,0.59,0.00}{\ul{\textbf{customer}}} as a role that specifically characterises \textcolor[rgb]{0.00,0.59,0.00}{\ul{\textbf{ordered product}}} role by their involvement in the activity, while \textcolor[rgb]{0.00,0.07,1.00}{\emph{receives}} represents the verb concept of the factual relationship between the two terms. Similarly, the Fact Type, ' \textcolor[rgb]{0.00,0.59,0.00}{\ul{\textbf{name}}} \textcolor[rgb]{0.00,0.07,1.00}{\emph{is of}} \textcolor[rgb]{0.00,0.59,0.00}{\ul{\textbf{customer}}}' shows \textcolor[rgb]{0.00,0.59,0.00}{\ul{\textbf{name}}} and \textcolor[rgb]{0.00,0.59,0.00}{\ul{\textbf{customer}}} as noun concepts, where \textcolor[rgb]{0.00,0.59,0.00}{\ul{\textbf{name}}} is an attribute to characterise \textcolor[rgb]{0.00,0.59,0.00}{\ul{\textbf{customer}}} while \textcolor[rgb]{0.00,0.07,1.00}{\emph{is of}} is a verb concept for the Fact Type.

The Terms and the Fact Types make up the {\it Business Vocabulary} in an SBVR model.

It might be worth noting that there is a synonymous form for Fact Types in SBVR \cite{WS04}, which allows to identify Fact Types that convey the same meaning. For example:\\
Fact Type: \textcolor[rgb]{0.00,0.59,0.00}{\ul{\textbf{name}}} \textcolor[rgb]{0.00,0.07,1.00}{\emph{is of}} \textcolor[rgb]{0.00,0.59,0.00}{\ul{\textbf{customer}}}; \qquad
Synonymous Form: \textcolor[rgb]{0.00,0.59,0.00}{\ul{\textbf{customer}}} \textcolor[rgb]{0.00,0.07,1.00}{\emph{has}} \textcolor[rgb]{0.00,0.59,0.00}{\ul{\textbf{name}}}

The synonymous form is useful when it comes to verifying the rule set in the SBVR model - whether all rules have been captured; whether any rules are in conflict.

\begin{figure}[!hbtp]
  \begin{center}
    \fbox{\includegraphics[width=0.985\textwidth]{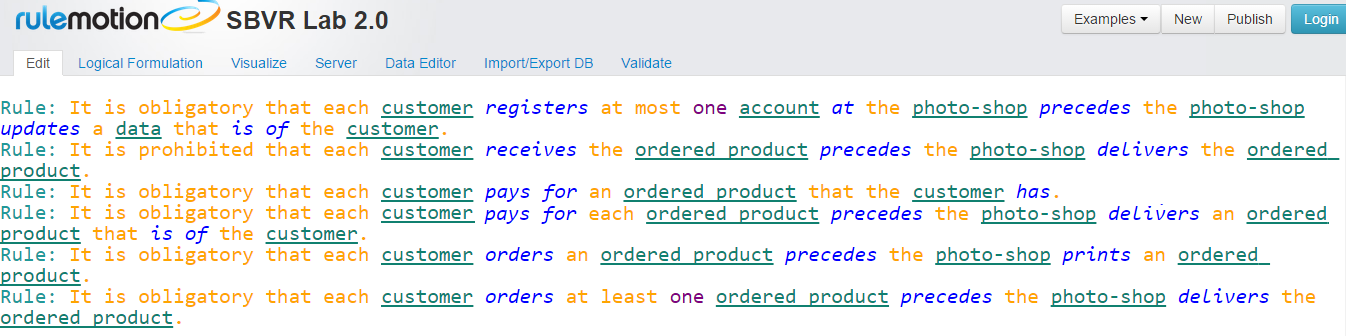}}
  \caption{Rules in the Online Photo Shop case study}
  \label{fig:Rules}
  \end{center}
\end{figure}

Fact Types are then used to construct the rules in the model as shown in Figure \ref{fig:Rules}. As mentioned in Section \ref{sec:SBVR}, rules are expressed using appropriate quantification, logical operations (if applicable) and modalities. In what follows we use the rules from Table \ref{table:ruleEx}.

\begin{table}[h]
\centering
  \begin{tabular}{|c|}
  \hline
  \textbf{Rule} \\
  \hline
    \textcolor[rgb]{1.00,0.50,0.00}{It is obligatory that} \textcolor[rgb]{1.00,0.50,0.00}{the}  \textcolor[rgb]{0.00,0.59,0.00}{\ul{\textbf{photo-shop}}}  \textcolor[rgb]{0.00,0.07,1.00} {\emph{delivers}}   \textcolor[rgb]{1.00,0.50,0.00}{the}  \textcolor[rgb]{0.00,0.59,0.00}{\ul{\textbf{ordered product}}} \textcolor[rgb]{1.00,0.50,0.00}{that} \textcolor[rgb]{0.00,0.07,1.00} {\emph{is of}} \textcolor[rgb]{1.00,0.50,0.00}{each}   \textcolor[rgb]{0.00,0.59,0.00}{\ul{\textbf{customer}}}\\ \textcolor[rgb]{1.00,0.50,0.00}{and at least one}  \textcolor[rgb]{0.00,0.59,0.00}{\ul{\textbf{ordered product}}} \textcolor[rgb]{1.00,0.50,0.00}{that} \textcolor[rgb]{0.00,0.07,1.00} {\emph{is of}} \textcolor[rgb]{1.00,0.50,0.00}{the}   \textcolor[rgb]{0.00,0.59,0.00}{\ul{\textbf{customer}}}\\
  \hline
\end{tabular}
  \caption{Business rules for the Online Photo Shop (the {\it Rules} in an SBVR model)}\label{table:ruleEx}
\end{table}

Semantic formulations are used in SBVR to structure the meaning of rules and come in two flavours; logical formulation and projection \cite{WS04}. All rules in the previous examples are structured according to the specialisations of logical formulation, e.g., logical operations, quantification, etc. An instantiation formulation also is one of the logical formulations that bind a concept (e.g., individual concept) to a bindable target (variable or an individual concept from logical formulation) to formulate the meaning. This is vital in the rules to express an accurate meaning.

For example, the specification in the case study (Table \ref{table:amended constraints}) attempts to prevent the situation where the photo shop attempts to deliver a product that no customer has ordered. To this effect, the logical formulation: '\textcolor[rgb]{1.00,0.50,0.00}{It is obligatory that} \textcolor[rgb]{1.00,0.50,0.00}{the}  \textcolor[rgb]{0.00,0.59,0.00}{\ul{\textbf{photo-shop}}}  \textcolor[rgb]{0.00,0.07,1.00} {\emph{delivers}}   \textcolor[rgb]{1.00,0.50,0.00}{the}  \textcolor[rgb]{0.00,0.59,0.00}{\ul{\textbf{ordered product}}} \textcolor[rgb]{1.00,0.50,0.00}{that} \textcolor[rgb]{0.00,0.07,1.00} {\emph{is of}} \textcolor[rgb]{1.00,0.50,0.00}{each}   \textcolor[rgb]{0.00,0.59,0.00}{\ul{\textbf{customer}}}' has been added to the set of rules in the SBVR model. According to this rule, the ordered product (photo/poster/album) which is a concept (delivered by the photo shop) binds to the ordered product which is an individual concept (ordered by the customer). Additionally, \textcolor[rgb]{1.00,0.50,0.00}{that}  which is located after the designation for a noun concept, \textcolor[rgb]{0.00,0.59,0.00}{\ul{\textbf{ordered product}}} here, and before the designation for verb concept, \textcolor[rgb]{0.00,0.07,1.00} {\emph{is of}}, is used to restrict \textcolor[rgb]{1.00,0.50,0.00}{that} keyword on the previous designation based on facts about them.

Moreover, Table \ref{table:ruleEx} shows the rule that is supposed to represent the ordering of the activities. The rule however does not capture the dependency between the activities, i.e., that activity 'deliver' by the photo shop must occur after activity '€œorder' by the customer of at least one product. Thus, this is why a notion of ordering between fact types in SBVR is proposed, in the next section, to capture the type of rules which prescribes the ordering of the underlying service interactions in a business scenario.

\section{A notion of time in business rules: ordering of service interactions}

We have seen that with respect to coordination it is necessary to capture the ordering of service interactions in order to encapsulate the important  properties of domain in a choreography. In the Introduction, we outlined some reasons why the ordering of service interactions is important. Without a notion of time, e.g., precedence, it is not straightforward to construct a rule which prohibits certain anomalies coming into view such as race conditions. For instance, the temporal ordering "precedence" need to be placed in between the Fact Type: \textcolor[rgb]{0.00,0.59,0.00}{\ul{\textbf{customer}}} \textcolor[rgb]{0.00,0.07,1.00}{\emph{pays for}} \textcolor[rgb]{0.00,0.59,0.00}{\ul{\textbf{ordered product}}}, and the Fact Type: \textcolor[rgb]{0.00,0.59,0.00}{\ul{\textbf{ordered product}}} \textcolor[rgb]{0.00,0.07,1.00} {\emph{is delivered by}}   \textcolor[rgb]{0.00,0.59,0.00}{\ul{\textbf{photo-shop}}}. In view of that example, it seems appropriate to look into expressing ordering of Fact Types in SBVR without changing the OMG standard or introducing special primitives particular to our approach. The idea is to express ordering in terms of dependency between certain messages (\emph{causality}), and by implication also choice ({\it conflict€}) and \emph{concurrency}, which should be possible, especially if a true concurrency semantics is pursued as done in \cite{S01}.


It transpires that such a notion of time, i.e., the ordering of service interactions in a choreography, is closely related to Type 2 in the {\it Date-Time Vocabulary} (DTV) \cite{WS08}, which is a supplementary specification to SBVR by OMG. Type 2 of Time aspects in DTV concerns a relationship between situation kinds and occurrences. The construction based on Type 2 in DTV (pp. 183-215 in \cite{WS08}) draws upon the concept of "state of affairs" in SBVR which refers to an event, activity, situation, or circumstance that is {\it actual} (in fact, defined as an {\it actuality} in SBVR spec \cite{WS04}). The actuality itself is an instance of a verb concept. For example, the proposition '\textcolor[rgb]{1.00,0.50,0.00}{a} \textcolor[rgb]{0.00,0.59,0.00}{\ul{\textbf{customer}}} \textcolor[rgb]{0.00,0.07,1.00}{\emph{pays for}} \textcolor[rgb]{1.00,0.50,0.00}{an}   \textcolor[rgb]{0.00,0.59,0.00}{\ul{\textbf{ordered product}}}' has the actuality (state of affairs), '\textcolor[rgb]{1.00,0.50,0.00}{an}   \textcolor[rgb]{0.00,0.59,0.00}{\ul{\textbf{ordered product payment}}}  \textcolor[rgb]{1.00,0.50,0.00}{that} \textcolor[rgb]{0.00,0.07,1.00} {\emph{is of}} \textcolor[rgb]{1.00,0.50,0.00}{a} \textcolor[rgb]{0.00,0.59,0.00}{\ul{\textbf{customer}}}' which is formed out of the verb concept \textcolor[rgb]{0.00,0.07,1.00}{\emph{pays for}} while both the proposition and state of affairs satisfy the subclause in SBVR spec \cite{WS04} which say that "\textcolor[rgb]{1.00,0.50,0.00}{it is necessary that each} \textcolor[rgb]{0.00,0.59,0.00}{\ul{\textbf{proposition }}} \textcolor[rgb]{0.00,0.07,1.00} {\emph{corresponds to}} \textcolor[rgb]{1.00,0.50,0.00}{exactly one} \textcolor[rgb]{0.00,0.59,0.00}{\ul{\textbf{state of affairs}}}".

However, DTV does not agree with that necessity because the proposition that corresponds to a state of affairs, as stated in the example, refers to one event only. On the contrary, the objective of DTV is to represent real states of affairs that occur multiple times. For this reason, DTV introduces a "situation kind" in place of state of affairs. A situation kind refers to a type of situation, event or activity that may occur multiple times. It is related closely with occurrence, so a typical example of a situation kind that occurs in actual situation at some place and time. For instance, the situation kind, '\textcolor[rgb]{1.00,0.50,0.00}{an}   \textcolor[rgb]{0.00,0.59,0.00}{\ul{\textbf{ordered product payment}}}  \textcolor[rgb]{1.00,0.50,0.00}{that} \textcolor[rgb]{0.00,0.07,1.00} {\emph{is of}} \textcolor[rgb]{1.00,0.50,0.00}{a} \textcolor[rgb]{0.00,0.59,0.00}{\ul{\textbf{customer}}}' refers to the activity of (binding) a customer paying their ordered product, which may occur multiple times in future.

Therefore, we apply a temporal ordering of situation kinds which is specifically based on the template '\textcolor[rgb]{0.00,0.59,0.00}{\ul{\textbf{situation kind\tiny{1}}}} \textcolor[rgb]{0.00,0.07,1.00} {\emph{precedes}} \textcolor[rgb]{0.00,0.59,0.00}{\ul{\textbf{situation kind\tiny{2}}}}' that further defines in SBVR that '\textcolor[rgb]{1.00,0.50,0.00}{each}  \textcolor[rgb]{0.00,0.59,0.00}{\ul{\textbf{occurrence}}} \textcolor[rgb]{0.00,0.07,1.00} {\emph{of}} \textcolor[rgb]{0.00,0.59,0.00}{\ul{\textbf{situation kind\tiny{1}}}} \textcolor[rgb]{0.00,0.07,1.00} {\emph{precedes}} '\textcolor[rgb]{1.00,0.50,0.00}{each} \textcolor[rgb]{0.00,0.59,0.00}{\ul{\textbf{occurrence}}} \textcolor[rgb]{0.00,0.07,1.00} {\emph{of}} \textcolor[rgb]{0.00,0.59,0.00}{\ul{\textbf{situation kind\tiny{2}}}}'. This allows comparing the ordering of two situation kinds. The following rule makes use of this temporal ordering: '\textcolor[rgb]{1.00,0.50,0.00}{It is obligatory that each} \textcolor[rgb]{0.00,0.59,0.00}{\ul{\textbf{customer}}} \textcolor[rgb]{0.00,0.07,1.00} {\emph{pays for}} \textcolor[rgb]{1.00,0.50,0.00}{each} \textcolor[rgb]{0.00,0.59,0.00}{\ul{\textbf{ordered product}}} \textcolor[rgb]{1.00,0.50,0.00}{that} \textcolor[rgb]{0.00,0.07,1.00} {\emph{is of}} \textcolor[rgb]{1.00,0.50,0.00}{the} \textcolor[rgb]{0.00,0.59,0.00}{\ul{\textbf{customer}}} \textcolor[rgb]{0.00,0.07,1.00} {\emph{precedes}} \textcolor[rgb]{1.00,0.50,0.00}{that} \textcolor[rgb]{0.00,0.59,0.00}{\ul{\textbf{ordered product}}} \textcolor[rgb]{0.00,0.07,1.00} {\emph{is delivered by}}   \textcolor[rgb]{1.00,0.50,0.00}{the}  \textcolor[rgb]{0.00,0.59,0.00}{\ul{\textbf{photo-shop}}}'. This is in fact the rule in the SBVR model of the choreography that captures the precedence constraint this section opened with.

There is also a temporal ordering of occurrences in DTV \cite{WS08} which can prescribe '\textcolor[rgb]{0.00,0.59,0.00}{\ul{\textbf{occurrence\tiny{1}}}}  \textcolor[rgb]{0.00,0.07,1.00} {\emph{precedes}} \textcolor[rgb]{0.00,0.59,0.00}{\ul{\textbf{occurrence\tiny{2}}}}'. In this Type 2 of time aspects in DTV, the "occurrence" is defined as the occurrence interval (specific time interval). Hence, it is not directly related to the notion of time considered here.

Our approach to ordering Fact Types in SBVR includes the use of the construct of \emph{objectification} in SBVR \cite{WS04}. This verb concept is used to specialise the "state of affairs" which in turn specialises "situation kinds" and "occurrence". Objectification may fill verb concept roles that range over a "situation kind" as well as an "occurrence". For example, the verb concept objectification as state of affairs for '\textcolor[rgb]{1.00,0.50,0.00}{an}   \textcolor[rgb]{0.00,0.59,0.00}{\ul{\textbf{ordered product payment}}}  \textcolor[rgb]{1.00,0.50,0.00}{that} \textcolor[rgb]{0.00,0.07,1.00} {\emph{is of}} \textcolor[rgb]{1.00,0.50,0.00}{a} \textcolor[rgb]{0.00,0.59,0.00}{\ul{\textbf{customer}}}' is defined as '\textcolor[rgb]{1.00,0.50,0.00}{a}   \textcolor[rgb]{0.00,0.59,0.00}{\ul{\textbf{customer }}} \textcolor[rgb]{0.00,0.07,1.00}{\emph{pays for}} \textcolor[rgb]{1.00,0.50,0.00}{an} \textcolor[rgb]{0.00,0.59,0.00}{\ul{\textbf{ordered product}}}'. It may be used with the verb concept '\textcolor[rgb]{0.00,0.59,0.00}{\ul{\textbf{photo-shop}}} \textcolor[rgb]{0.00,0.07,1.00} {\emph{needs}} \textcolor[rgb]{0.00,0.59,0.00}{\ul{\textbf{situation kind}}}' and also with the verb concept '\textcolor[rgb]{0.00,0.59,0.00}{\ul{\textbf{photo-shop}}} \textcolor[rgb]{0.00,0.07,1.00} {\emph{records}} \textcolor[rgb]{0.00,0.59,0.00}{\ul{\textbf{occurrence}}}'.
Thus, to express the ordering of service interactions in a choreography, we use a notion similar to Type 2 in DTV \cite{WS08} and apply the {\it objectification} construct of SBVR \cite{WS04}. An example rule that uses this notion of time ordering from the Online Photo Shop is given by:\\
\\*
'\textcolor[rgb]{1.00,0.50,0.00}{It is obligatory that each} \textcolor[rgb]{0.00,0.59,0.00}{\ul{\textbf{customer }}} \textcolor[rgb]{0.00,0.07,1.00}{\emph{pays for}} \textcolor[rgb]{1.00,0.50,0.00}{each} \textcolor[rgb]{0.00,0.59,0.00}{\ul{\textbf{ordered product}}} \textcolor[rgb]{1.00,0.50,0.00}{that} \textcolor[rgb]{0.00,0.07,1.00} {\emph{is of}} \textcolor[rgb]{1.00,0.50,0.00}{a} \textcolor[rgb]{0.00,0.59,0.00}{\ul{\textbf{customer}}} \textcolor[rgb]{0.00,0.07,1.00}{\emph{precedes}} \textcolor[rgb]{1.00,0.50,0.00}{that} \textcolor[rgb]{0.00,0.59,0.00}{\ul{\textbf{ordered product}}} \textcolor[rgb]{0.00,0.07,1.00} {\emph{is delivered by}}   \textcolor[rgb]{1.00,0.50,0.00}{the}  \textcolor[rgb]{0.00,0.59,0.00}{\ul{\textbf{photo-shop}}}'\\
\\*
This rule has the following two propositions or Fact Types:
'\textcolor[rgb]{0.00,0.59,0.00}{\ul{\textbf{customer }}}
\textcolor[rgb]{0.00,0.07,1.00}{\emph{pays for}}
\textcolor[rgb]{0.00,0.59,0.00}{\ul{\textbf{ordered product}}}' and
'\textcolor[rgb]{0.00,0.59,0.00}{\ul{\textbf{ordered product}}}
\textcolor[rgb]{0.00,0.07,1.00} {\emph{is delivered by}}
\textcolor[rgb]{0.00,0.59,0.00}{\ul{\textbf{photo-shop}}}'. Therefore,
the situation kinds, '\textcolor[rgb]{1.00,0.50,0.00}{an}
\textcolor[rgb]{0.00,0.59,0.00}{\textbf{\ul{ordered product} \ul{payment}}}
\textcolor[rgb]{1.00,0.50,0.00}{that} \textcolor[rgb]{0.00,0.07,1.00} {\emph{is
of}} \textcolor[rgb]{1.00,0.50,0.00}{a}
\textcolor[rgb]{0.00,0.59,0.00}{\ul{\textbf{customer}}}' and
'\textcolor[rgb]{1.00,0.50,0.00}{the}
\textcolor[rgb]{0.00,0.59,0.00}{\ul{\textbf{ordered product delivery}}}
\textcolor[rgb]{0.00,0.07,1.00} {\emph{by}} \textcolor[rgb]{1.00,0.50,0.00}{the}
\textcolor[rgb]{0.00,0.59,0.00}{\ul{\textbf{photo-shop}}}' are an
actuality denoted by the verb concept objectification of the two propositions or
Fact Types.

\begin{figure}[!hbtp]
  \begin{center}
    \fbox{\includegraphics[width=0.63\textwidth]{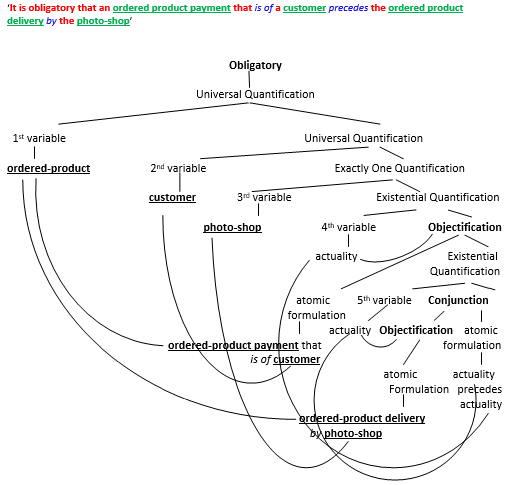}}
  \caption{ SBVR Logical Formulation}\label{lf}
  \end{center}
\end{figure}

By considering the vocabulary structure in SBVR: '\textcolor[rgb]{0.00,0.59,0.00}{\ul{\textbf{situation kind\tiny{1}}}} \textcolor[rgb]{0.00,0.07,1.00} {\emph{precedes}} \textcolor[rgb]{0.00,0.59,0.00}{\ul{\textbf{situation kind\tiny{2}}}}'  and taking '\textcolor[rgb]{1.00,0.50,0.00}{an} \textcolor[rgb]{0.00,0.59,0.00}{\ul{\textbf{ordered product payment}}}  \textcolor[rgb]{1.00,0.50,0.00}{that} \textcolor[rgb]{0.00,0.07,1.00} {\emph{is of}} \textcolor[rgb]{1.00,0.50,0.00}{a} \textcolor[rgb]{0.00,0.59,0.00}{\ul{\textbf{customer}}}' to be \textcolor[rgb]{0.00,0.59,0.00}{\ul{\textbf{situation kind\tiny{1}}}}, and '\textcolor[rgb]{1.00,0.50,0.00}{the} \textcolor[rgb]{0.00,0.59,0.00}{\ul{\textbf{ordered product delivery}}} \textcolor[rgb]{0.00,0.07,1.00} {\emph{by}} \textcolor[rgb]{1.00,0.50,0.00}{the} \textcolor[rgb]{0.00,0.59,0.00}{\ul{\textbf{photo-shop}}}'  to be \textcolor[rgb]{0.00,0.59,0.00}{\ul{\textbf{situation kind\tiny{2}}}}  the expression of the rule is given as:\\
\\*
'\textcolor[rgb]{1.00,0.50,0.00}{It is obligatory that an} \textcolor[rgb]{0.00,0.59,0.00}{\ul{\textbf{ordered product payment}}}  \textcolor[rgb]{1.00,0.50,0.00}{that} \textcolor[rgb]{0.00,0.07,1.00} {\emph{is of}} \textcolor[rgb]{1.00,0.50,0.00}{a} \textcolor[rgb]{0.00,0.59,0.00}{\ul{\textbf{customer}}} \textcolor[rgb]{0.00,0.07,1.00}{\emph{precedes}} \textcolor[rgb]{1.00,0.50,0.00}{the} \textcolor[rgb]{0.00,0.59,0.00}{\ul{\textbf{ordered product delivery}}} \textcolor[rgb]{0.00,0.07,1.00} {\emph{by}} \textcolor[rgb]{1.00,0.50,0.00}{the} \textcolor[rgb]{0.00,0.59,0.00}{\ul{\textbf{photo-shop}}}'\\
\\*
This ability to express temporal constraints in business rules has been applied to our case study to show ordering of service interactions. All the rules in a choreography are transformed to the SBVR Logical Formulation (Ch. 10 in \cite{WS04}) as shown in Figure \ref{lf}.

Figure \ref{lf} shows the representation of the Logical Formulation of the stated rule. It represents the relationships between the obligation formulation, the atomic formulations, the instantiation formulation, and the objectification in the rules based on the Simplified Syntax for Logical Formulations in Annex F of DTV \cite{WS08}. The instantiation formulation in the logical formulation can be seen in Figure \ref{lf} whenever the first variable '\textcolor[rgb]{0.00,0.59,0.00}{\ul{\textbf{ordered product}}}' binds to the concepts '\textcolor[rgb]{0.00,0.59,0.00}{\ul{\textbf{ordered product payment}}}' and '\textcolor[rgb]{0.00,0.59,0.00}{\ul{\textbf{ordered product delivery}}}' to show both concepts are referring to the same ordered product that was ordered and paid for by the customer. Then, the rules are translated into first order logic, following the SBVR Logical Formulation \cite{WS04}. This transformation opens up the space for reasoning (verification) as well as model transformations between different tools.


An example of the translation for one rule from our case study is given in Table \ref{table:FOL}. The first order logic expression $o\forall{x}\forall{y}\exists^1{z}(B(P(y,x)\wedge Q(x,y),T(x,z)))$ says that "for all products, $x$, for all customers, $y$, there exists at most one photo shop, $z$, such that $y$ pays for $x$, and $x$ is of $y$, precedes $x$ is delivered by $z$".
This is half way between the logic and the structured English of the business rule.

\begin{table}[!hbtp]
  \centering
  \begin{tabular}{|l|l|l|}
    \hline
     \textbf{Declaration} & \textbf{Business Rule} & \textbf{First Order Logic} \\
     \hline
    $x$ is an ordered product; & \textcolor[rgb]{1.00,0.50,0.00}{It is obligatory that each} & $o\forall{x}\forall{y}\exists^1{z}(B(P(y,x)\wedge Q(x,y),T(x,z)))$  \\
    $y$ is a customer; &  \textcolor[rgb]{0.00,0.59,0.00}{\ul{\textbf{customer }}} \textcolor[rgb]{0.00,0.07,1.00}{\emph{pays for}} \textcolor[rgb]{1.00,0.50,0.00}{each} & \\
    $z$ is a photo-shop; & \textcolor[rgb]{0.00,0.59,0.00}{\ul{\textbf{ordered product}}} \textcolor[rgb]{1.00,0.50,0.00}{that} \textcolor[rgb]{0.00,0.07,1.00} {\emph{is of}} & \\
    $P(y,x):y$ pays for $x$; & \textcolor[rgb]{1.00,0.50,0.00}{a} \textcolor[rgb]{0.00,0.59,0.00}{\ul{\textbf{customer}}} \textcolor[rgb]{0.00,0.07,1.00} {\emph{precedes}} \textcolor[rgb]{1.00,0.50,0.00}{that} & \\
    $T(x,z):x$ is delivered by $z$; & \textcolor[rgb]{0.00,0.59,0.00}{\ul{\textbf{ordered product}}} &\\
    $Q(x,y):x$ is of $y$; & \textcolor[rgb]{0.00,0.07,1.00} {\emph{is delivered by}} \textcolor[rgb]{1.00,0.50,0.00}{the}  & \\
    $B(P(y,x)\wedge Q(x,y),T(x,z))$: & \textcolor[rgb]{0.00,0.59,0.00}{\ul{\textbf{photo-shop}}}   &  \\
    $P(y,x)\wedge Q(x,y)$ precedes $T(x,z)$ &   & \\
    \hline
  \end{tabular}
    \caption{SBVR First Order Logic}\label{table:FOL}
\end{table}


\section{Related Work}
\label{sec:relatedwork}


The work on \emph{DecSerFlow} proposed in \cite{M10} provides a declarative language together with a logical framework for reasoning while the work described in \cite{H01} uses the implementation of the {\it Business Process Execution Language for Web Services} (BPEL4WS) (see the specification document \cite{WS03}) and its semantics for choreography specification. The representation of a choreography is given in the form of graphical specification of service flows which can be mapped onto Linear Temporal Logic (LTL) \cite{H07} and in the form of XML data format definition which is then translated to Finite State Process (FSP) process algebra, respectively, thus allowing to model the required behaviour. This enriches the expressiveness and allows to perform interoperability and verification tasks, including conformance checking and deadlock detection \cite{M02}. \cite{H01} also provides a tool, LTSA-WS for checking the correctness of the service interactions in terms of whether they correspond to those specified in the requirements. While these approaches provide reasoning capability, \emph{DecSerFlow} is a proprietary language while \cite{H01} uses both an informal and a formal language that require training to read and write for specifying software services. Hence, both comprise a learning curve for practitioners (business analysts or stakeholders). 

Furthermore, \cite{Sc02} also place emphasis on coordination of service interactions that correlate to the choreography specification - a model derived from BPMN 2.0 \cite{WS05} is implemented. On the other hand, \cite{M03} provides a declarative approach and applies UML activity diagram (as illustration purposes) for describing and capturing the ordering constraints between interactions, yet for service interface adaptation. Additionally, \cite{C01} provides an integrated tool support for the specification and validation and verification for adaptation contracts. In other words, \cite{M03} and \cite{C01} focus on discovering mismatches between behavioural interfaces which this different with our focus as discussed earlier in Section \ref{sec:intro}.

The approach built around SBVR \cite{WS04} which we propose in this paper for modelling service choreographies uses a structured natural language standard maintained by OMG, which comes with a logical formulation (recall Section 4) which can be exploited for reasoning about correctness. This means that an SBVR model can be parsed and is machine readable. We have more to say on this in the concluding section of the paper (Section \ref{sec:concl}).

In introducing a notion of time, understood in terms of ordering of Fact Types, we chose not to come from the angle of the \emph{Object Constraint Language} (OCL) \cite{WS09} or Allen's temporal logic \cite{AF94}. This is because although these approaches deal with time, OCL introduces Collections to manage an \emph{OrderedSet} and a \emph{Sequence} which uses elements to represent the occurrence of objects. Both constructs contain a collection where the elements are ordered. However, the \emph{OrderedSet} contains unique elements which means no duplicates of the elements exist while the elements in Sequence may be part of a sequence more than once. For example, consider a class Employee with an attribute 'age'. Collection contains three employees such as \emph{employee1.age = 24}, \emph{employee2.age = 27} and \emph{employee3.age = 27}. Thus,\\
\textbf{Expression} : self.employees $\rightarrow$ sortedBy(age);\\
\textbf{Result} : Sequence: {\it employee1, employee2, employee3}

In \cite{AF94} a temporal logic was developed to represent knowledge of properties, events, and actions using one primitive object, namely the time period, and one primitive relation €'œMeets'€ (\emph{m} and \emph{n} meet if and only if \emph{m} precedes \emph{n}). One can describe \emph{m} precedes \emph{n}, but \emph{m} and \emph{n} represent two time periods. Hence, both OCL and Allen's temporal logic are not suitable to address the ordering of Fact Types inside a rule.

It might be instructive to also compare the way a choreography can be expressed in {\it DecSerFlow} \cite{M10} and in SBVR. For example, \emph{succession(A,B)} is used in \emph{DecSerFlow} to constrain the Online Photo Shop rules: the shop will not 'open order'€ before the customer executes activity '€œregister'€. Hence, the 'succession' constraint replaces parameter '€œA'€ with activity '€œregister' and parameter '€œB'€ with activity 'open order'€. It is then converted to a logic expression (in Linear Temporal Logic (LTL)) in the following DecSerFlow template:\\
\\*
\emph{succession(register, open order) = response(register, open order) precedence(register, open order)}\\
\\*
In contrast, in SBVR the corresponding rule will be\\

€œ\textcolor[rgb]{1.00,0.50,0.00}{It is obligatory that each} \textcolor[rgb]{0.00,0.59,0.00}{\ul{\textbf{customer }}} \textcolor[rgb]{0.00,0.07,1.00}{\emph{registers}} \textcolor[rgb]{1.00,0.50,0.00}{at most one} \textcolor[rgb]{0.00,0.59,0.00}{\ul{\textbf{account}}} \textcolor[rgb]{0.00,0.07,1.00}{\emph{at}} \textcolor[rgb]{1.00,0.50,0.00}{the} \textcolor[rgb]{0.00,0.59,0.00}{\ul{\textbf{photo-shop}}} \textcolor[rgb]{0.00,0.07,1.00}{\emph{precedes}} \textcolor[rgb]{1.00,0.50,0.00}{the} \textcolor[rgb]{0.00,0.59,0.00}{\ul{\textbf{photo-shop}}} \textcolor[rgb]{0.00,0.07,1.00}{\emph{updates}} \textcolor[rgb]{1.00,0.50,0.00}{the} \textcolor[rgb]{0.00,0.59,0.00}{\ul{\textbf{data}}}
\textcolor[rgb]{1.00,0.50,0.00}{that} \textcolor[rgb]{0.00,0.07,1.00}{\emph{is of}} \textcolor[rgb]{1.00,0.50,0.00}{the} \textcolor[rgb]{0.00,0.59,0.00}{\ul{\textbf{customer }}}.\\

We believe that this global constraint is more understandable to domain specialists (business analysts) but also humans in general.

\section{Conclusion and Future Work}
\label{sec:concl}

In this paper, we presented a declarative approach to the coordination of distributed applications comprising stand-alone services. We proposed the use of SBVR for choreography specification and demonstrated how \emph{objectification} and \emph{actuality} can be exploited in impressing temporal aspects between Fact Types that appear in a rule. This notion of time ordering is reminiscent of Type 2 in DTV \cite{WS08}, a supplementary specification document to SBVR, also advocated by OMG, and is useful in capturing the global constraints in a multi-party conversation involved in  a service choreography.

SBVR can be used to support the development of ontologies and business models using structured natural language. It is widely used to cope with complex requirements of business operations with a language that is easily understandable by business analysts (domain specialists) rather than systems analysts (implementation specialists). We are currently exploring the possibility of integrating SBVR with the work on participatory modelling, where we use Fuzzy Cognitive Maps (FCM) and techniques from network analytics in identifying strategic intervention points in complex networks. This is looking into the Humber region, UK, (one of the UK's most important energy hubs) as a case study where local authorities and various groups of stakeholders engage in moving from a fossil fuel economy to a bio-based economy \cite{SHBchapter}. The FCM map is built over a one-day workshop with stakeholders and the whole process could be helped by using SBVR to capture expertise and dependencies in the complex network of the Humber region in a way that is understandable to business, policy makers and researchers alike.

As discussed in Section 4, the SBVR Logic Formulation prescribed in the OMG specification document for SBVR \cite{WS04} can be used to transform an SBVR model into first order logic, which can be useful for reasoning by looking at the \emph{FODAL} approach that proposed by \cite{D01} as discussed in Section 2. In the context of coordination, and service choreography, temporal aspects will need to be handled and possibly in a distributed manner as done in Mdtl \cite{FM07} which we have used for reasoning about distributed and concurrent interactions before, or as done for decentralized self-adaptive systems in \cite{LT01} where behavioural properties are specified using timed computation tree logic (TCTL). Hence, with respect to choreography verification we would typically be looking at Computational Tree Logic (CTL) or Linear Temporal Logic (LTL) \cite{H07} . LTL models time as a sequence of states, extending infinitely into the future. However, it does not allow to quantify explicitly over paths. CTL allows to reason about sequences of events that capture the semantics. The CTL syntax includes a parse tree, a quantifier equivalent to logical formulation kinds, and both CTL and LTL syntaxes denote a set of atomic propositions which is similar to the first order logic proposition used in the logical foundation for the SBVR model here.



\begin{figure}[!hbtp]
 \begin{center}
    \fbox{\includegraphics[width=0.8\textwidth]{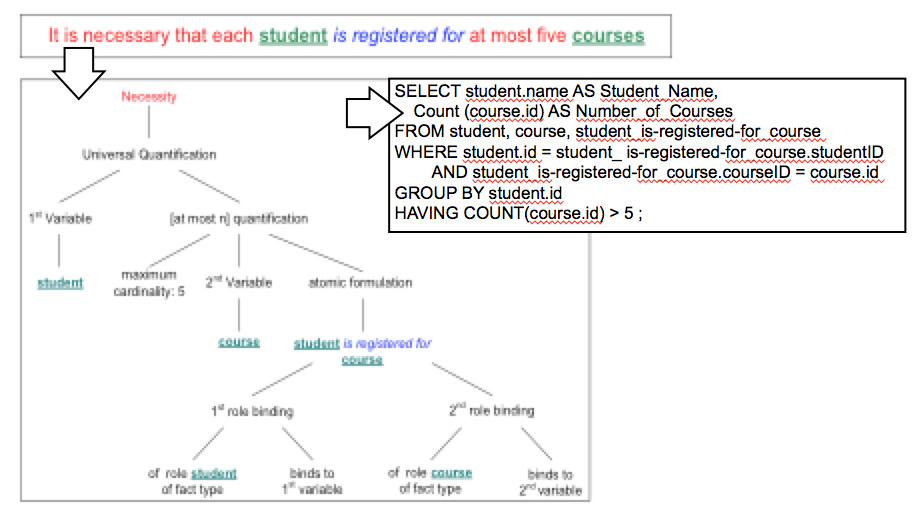}}
  \caption{SBVR rules transformed to SQL queries \cite{MMK10}}\label{fig:sbvr2sql}
  \end{center}
\end{figure}

In terms of implementation, apart from the SBVR editor \cite{SBVReditor} discussed earlier (Section 2) we also implemented an SBVR to SQL comipler \cite{SBVR2SQL}. Using the Logical Formulation SBVR-LF \cite{WS02}, the complier maps business rules onto SQL queries to be executed on a (relational) database, as shown in Figure \ref{fig:sbvr2sql}. The relational database is also automatically generated by the model.
The work in \cite{M01} demonstrates how information systems can be generated directly from SBVR.
Therefore, SBVR can be used to formulate complex data queries in a way that provides a higher level of abstraction than SQL (or any other query language) as shown in \cite{MMK10}.

In previous work we have used {\it vector languages}, a model of true concurrency, to equip service choreographies with transactional guarantees, in the so-called {\it transaction languages} \cite{S01}. A natural extension of this work is to investigate the use of transaction languages, which is tuple-based formalism that captures the ordering of observable actions in a given choreography, to activate rules in the overlaid SBVR model of the service choreography. So transaction languages effectively play the role of the {\it blackboard}, in the sense of the work in \cite{RB01}, while the SBVR rules which are amenable to business analysts are used to constrain the generated implementation and seamlessly force it to adhere to the choreography specification \cite{Sc02}.

One other possible direction for future work then will focus on analysing the complete set of behaviours (all possible outcomes) by exploiting the logical underpinning of the SBVR model of the choreography. This would target reasoning and choreography verification tasks such as {\it realisation} and {\it conformance}.  In addition, another possible future extension has to do with the (correctness of the) transformation from natural language to SBVR \cite{Bordbar11}. Controlled Language (CL) is used in \cite{LN13} to ensure correctness and deadlock-freedom. This would allow our approach to extend to natural language for specifying choreographies and would appeal to a wider business audience.

\bibliographystyle{eptcs}
\bibliography{sbvr}

\end{document}